\documentclass{aa}
\newcommand{\be} {\begin{equation}}

\newcommand{\ee} {\end{equation}}
\newcommand{\etal}{ et al. }
\newcommand{\src}{4U\thinspace0142+61}
\newcommand{\R}{{\em ROSAT}}
\newcommand{\E}{{\em Einstein}}
\newcommand{\BSAX}{{\em Beppo}SAX} 
\newcommand{\RXTE}{RXTE}
\newcommand{\bc}{\begin{center}}
\newcommand{\ec}{\end{center}}
\newcommand {\rchisq}{$\chi_{\nu} ^{2}$}
\newcommand {\chisq}{$\chi^{2}$}
\newcommand {\rc}{\rm}
\def \nh {N${\rm _H}$}
\def \hcm {\hbox {\ifmmode $ atoms cm$^{-2}\else atoms cm$^{-2}$\fi}}
\input psfig.tex
\begin{document}

\thesaurus{({\rc 08.05.3;} 08.14.1; 08.16.7; 13.25.5)}
\title{\BSAX\ monitoring of the ``anomalous'' X--ray pulsar \src}
\author{G.L. Israel\inst{1, }\thanks{Affiliated to I.C.R.A.} \and 
T. Oosterbroek\inst{2} \and L.~Angelini\inst{3, }\thanks{Universities 
Space Research Association} \and S. Campana\inst{4}$^{, \star}$ 
\and S. Mereghetti\inst{5} \and A.N. Parmar\inst{2} \and A. Segreto\inst{6} 
\and L.~Stella\inst{1}$^{, \star}$ \and J.~van Paradijs\inst{7,8} 
\and N.E.~White\inst{3}} 

\institute{
Osservatorio Astronomico di Roma, Via Frascati 33, I--00040, 
Monteporzio Catone (Roma), Italy
\and 
Astrophysics Division, Space Science Department of ESA, 
ESTEC, P.O. Box 299, NL 2200 AG Noordwijk, The Netherlands
\and
Laboratory for High Energy Astrophysics, Code 662, 
NASA -- Goddard Space Flight Center, Greenbelt, MD 20771, USA
\and
Osservatorio Astronomico di Brera, Via E. Bianchi 46, 
I--23807  Merate (Lecco), Italy
\and
Istituto di Fisica Cosmica ``G. Occhialini", CNR, Via Bassini 15, I--20133 Milano, Italy
\and
Istituto di Fisica Cosmica ed Applicazioni all'Informatica, CNR, Via U. La
Malfa 153, I--90146 Palermo, Italy
\and 
Astronomical Institute ``Anton Pannekoek'' \& Center for 
High--Energy Astrophysics, Kruislaan 403, NL 1098 SJ Amsterdam, 
The Netherlands
\and 
Physics Department, UAH, Huntsville, AL 35899, USA 
}
\date{Received 25 January 1999 / Accepted 29 March 1999}
\offprints{G.L.~Israel (israel@coma.mporzio .astro.it)}
\maketitle
\markboth{Israel \etal\ : \BSAX\ monitoring of the ``anomalous'' X--ray 
                        pulsar \src}
         {Israel \etal\ : \BSAX\ monitoring of the ``anomalous'' X--ray 
                        pulsar \src}

\begin{abstract}
The 8.7\,s X--ray pulsar \src\ was monitored by \BSAX\ between January 
1997 and February 1998. This source belongs to the rapidly growing 
class of ``anomalous" X--ray pulsars (AXPs) which have pulse periods in the 
6--12\,s range 
and no plausible optical, infrared or radio counterparts. The \BSAX\ periods 
measurements {\rc show} that \src\ continues its spin--down at a nearly constant 
rate of \.P\,$\sim$\,2$\times$10$^{-12}$ s s$^{-1}$. The 0.5--10 keV pulse shape 
is double peaked. 
The phase--averaged spectrum can be described by a 
steep absorbed power--law ($\Gamma$ $\sim$ 4) plus a blackbody with  
$kT$ $\sim$ 0.4 keV. 

\src\ was also {\rc detected serendipitously} in a March 1996 \RXTE\ observation 
pointed towards the nearby 1455\,s X--ray pulsar RX\,J0146.9+6121. 
The timing analysis results are reported for this observation and the 
spin  history of \src\ since 1979 is discussed.    
\end{abstract}


\section{Introduction}

The properties of \src\ (White \etal 1987) remained puzzling for a 
long time, owing to confusion with a nearby pulsating and 
transient Be/neutron star system RX\,J0146.9+6121 (Motch \etal 
1991; Mereghetti {\rc \etal} 1993). 
The 1--10~keV spectrum is extremely soft (power law photon index of 
$\sim 4$, White \etal 1987) and led to the initial classification of 
\src\ as a possible black hole candidate. 
ASCA observations provide evidence for a $\sim 0.4$~keV 
blackbody component contributing $\sim 40$\% of the 
0.5--10 keV band X--ray flux (White \etal 1996). The X--ray luminosity of 
4U~0142+61 has not shown substantial secular variations around an average 
value of $\sim~6 \times 10^{34}$~erg~s$^{-1}$ (assuming a distance of 
1~kpc). Despite the small error box 
(5\arcsec\ radius), no optical or IR counterpart has yet been identified, 
down to $V<24$, $R<22.5$, $J<20$ and $K<17$ (Steinle \etal 1987; White \etal 1987; 
Coe \& Pightling 1998). These limits rule out the presence of a massive companion.   
Using data from the EXOSAT archive, Israel \etal (1994) 
discovered pulsations at 8.7 s, which were later confirmed 
with ROSAT (Hellier 1994). No delays in the pulse arrival times caused by 
orbital motion were found, with upper limits on $a_{\rm x}$\,sin $i$  of 
about $\sim 0.37$~lt--s for 
orbital periods $P_{\rm orb}$ between 7~min and 12~hr (Israel \etal 1994). 
{\rc Tighter} upper limits on the $a_{\rm x}\sin i$ ($\sim 0.26$~lt--s 
for 70\,s\,$\leq P_{\rm orb} \leq$ 2.5\,days) have been recently obtained with a \RXTE\ 
observation (Wilson \etal 1998). This yielded strong constraints on the orbital 
inclination and the mass of the possible companion star in the case of normal or  
helium main sequence star and giants with helium core. 
A white dwarf companion would be compatible both with current optical photometric 
and pulse arrival time limits.  
\begin{table*}
\caption[]{\src\ \BSAX\ Observation log}
\begin{tabular}{cccccccc}
\hline 
Start Time & Stop Time & MECS T$_{\rm exp}$ & Active MECSs & Count Rate LECS & 
MECS & Off--axis & Obs \\
 &  & s & \# & c/s & c/s & \arcmin\ & \\ 
\hline
%28--Mar--96~~11:17 & 28--Mar~~22:09 & ????? &   ----       & ---- & 24 & --- \\ 
97 Jan 03~~05:47 & Jan 04~~02:09 & 48226 &3& 1.03$\pm$0.01& 
1.70$\pm$0.01 &5 & A\\
97 Aug 09~~23:13 & Aug 10~~07:38 & 16757 &2& 1.01$\pm$0.02 & 
1.47$\pm$0.02&3 & B\\
98 Jan 26~~12:13 & Jan 27~~00:23 & 21785 &2& 
(4.2$\pm$0.4)$\times$10$^{-2}$&0.47$\pm$0.01 &20 & C\\
98 Feb 03~~09:11 & Feb 04~~02:37 & 31150 &2& 0.97$\pm$0.02& 
1.36$\pm$0.01&5 & D\\ 
\hline
\hline
\end{tabular}
\\ ~ \\ Note --- Count rates are not vignetting corrected. Vignetting is a factor 
1.10, 1.20 and 3.0 for an off--axis angle of 3, 5 and 20 arcmin, rescpectively. 
\end{table*}
%FIGURE 1
\begin{figure*}[tbh]
\centerline{
\psfig{figure=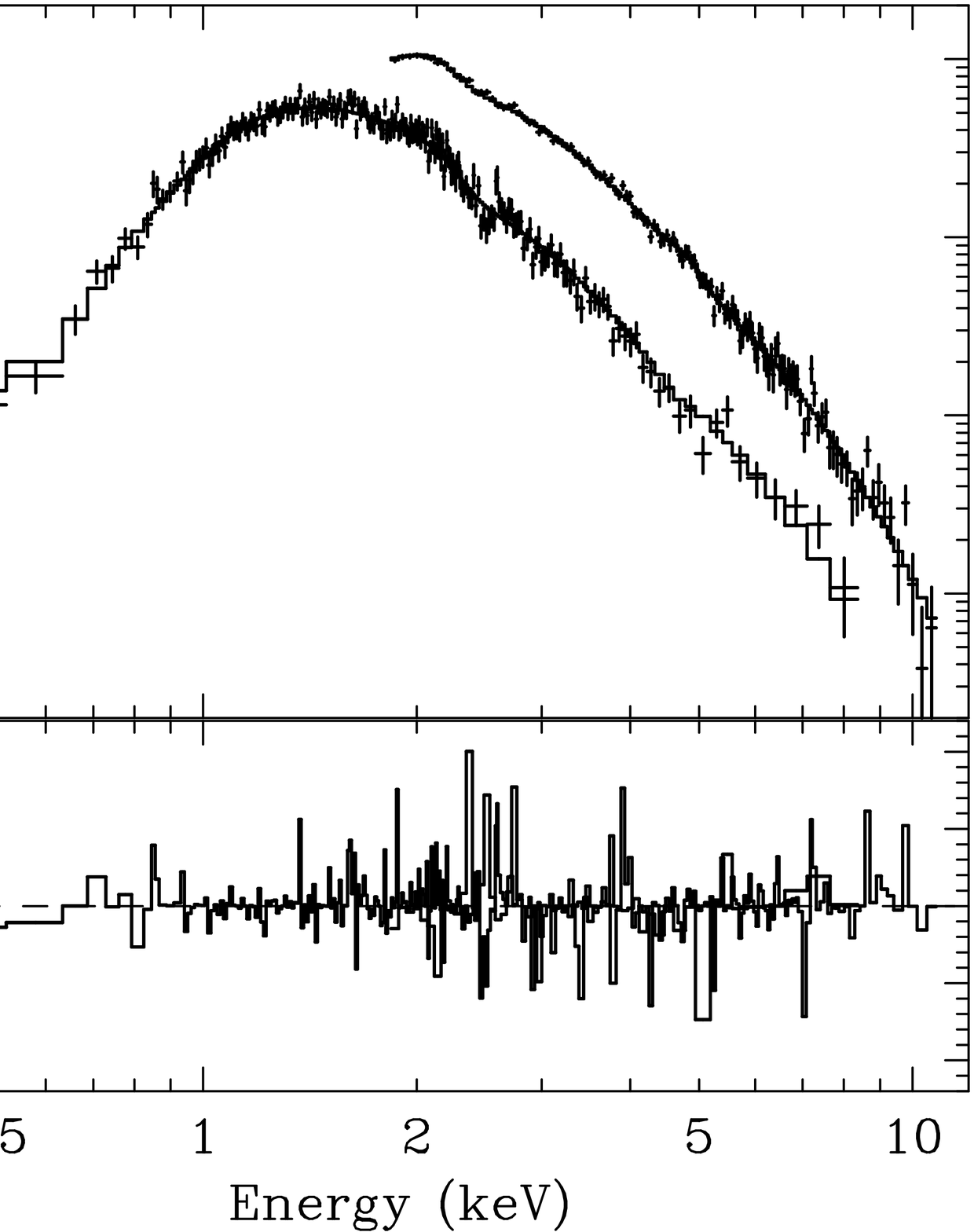,width=6cm,height=8cm}
\psfig{figure=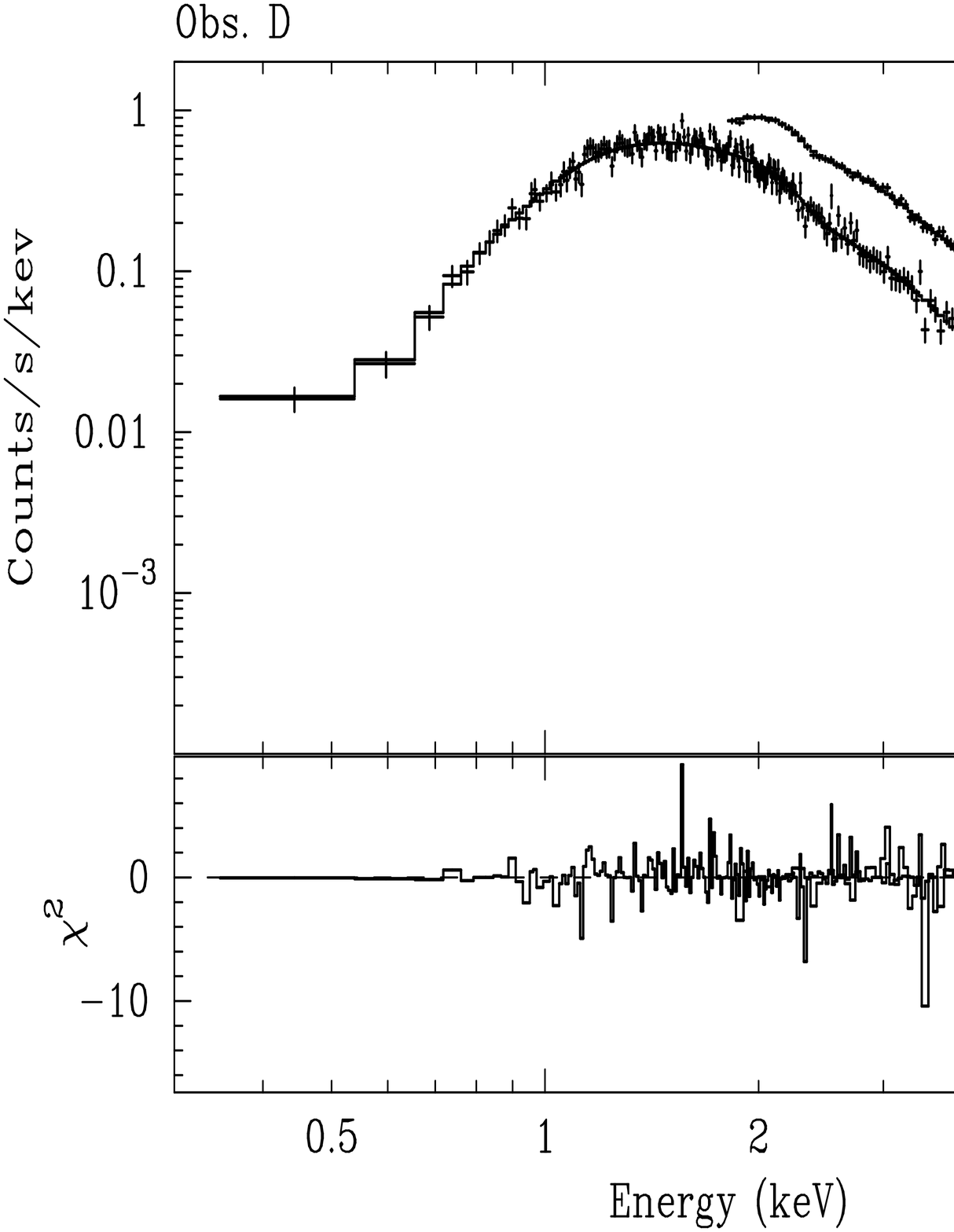,width=6cm,height=8cm}}
\caption{LECS and MECS energy spectra of \src\ during observation A (left 
panel) 
and D (right panel). The best fit spectral model (PL+BB; see text for 
details) 
is also shown together with the \chisq\ residuals.}
\end{figure*}
The EXOSAT and ROSAT period measurements, obtained in 1984 and 1993, provide a 
spin--down rate of $\sim 2.1\times10^{-12}$~s~s$^{-1}$.

The properties of \src\ are similar to those of a small group   
of ``anomalous'' X--ray pulsars (AXPs), with spin 
periods within a narrow range (6--12\,s; Mereghetti \& Stella 1995).  
Among these are 1E\,2259+586, 1E\,1048.1--5937, 1E\,1841--045 
(Vasisht \& Gotthelf 1997), 1RXS\,170849--400910 (Sugizaki \etal 1997) and 
AX\,J1845--045 (Torii \etal 1998; Gotthelf \& Vasisht 1998). 
   
We present here a detailed analysis of the \BSAX\ Narrow Field Instruments (NFIs) 
observations of the AXP \src. We confirm the presence of a blackbody 
spectral component in the soft X--ray spectrum as seen in the ASCA data 
(White \etal 1996) and in other two ``anomalous'' X--ray pulsars: 
1E\,2259+589 (Corbet \etal 1995; Parmar \etal 1998) and 1E\,1048.1--5937 
(Oosterbroek \etal 1998). We also present the results of pulse 
phase spectroscopy and the pulse period history of \src. {\rc We also discuss  
the timing analysis from a serendipitous Rossi X--ray Timing Explorer 
(\RXTE) observation of \src.}\\

\section{\BSAX\ Observations} 

Results from the Low--Energy Concentrator Spectrometer 
(LECS; 0.1--10~keV; Parmar et al. 1997) 
and Medium--Energy Concentrator Spectrometer (MECS; 1.3--10~keV; 
Boella et al. 1997) on--board \BSAX\ are presented.
{\rc The MECS consists of three identical grazing incidence telescopes
with imaging gas scintillation proportional counters in their focal planes.
The LECS uses an identical concentrator system as the MECS, but utilizes an
ultra-thin (1.25~$\mu$m) detector entrance window and a driftless 
configuration to extend the low-energy response to 0.1~keV. The fields of view
(FOV) of the LECS and MECS are circular with diameters of 37\arcmin\ and 
56\arcmin\, respectively. The energy resolution of both instruments is
$\simeq$ $8.5\sqrt{(6keV/{\rm E})}$ \% full--width half maximum
(FWHM), where E is the energy.} 
%FIGURE 2
\begin{figure}[tbh]
\centerline{\psfig{figure=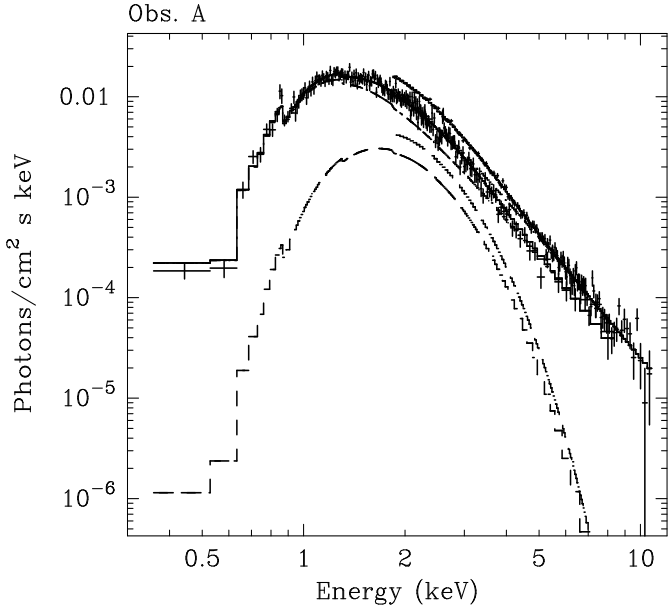,width=6cm,height=8cm}}
\caption{\src\ LECS and MECS unfolded spectrum for observation A. The 
power--law 
and blackbody components are shown (dotted and stepped lines).}
\end{figure}

\src\ was observed by \BSAX\ four times between January 1997 and 
February 1998 (see Table\,1). One of the goals of the program 
was to monitor the possible X--ray flux variations on a 
time scale of months with two 40\,ks pointings. {\rc However} between the 
first and the subsequent observations one of the three MECS units failed (1997 May 9), 
and data of observations B, C and D were obtained with the remaining two 
MECS units. Moreover during the second and third observations \BSAX\ experienced 
pointing failures resulting in a shorter effective exposure time.

\subsection{Spectral analysis}

Spectra were obtained centered on the position of \src\ using  an extraction 
radius of 8\arcmin\   for both  the LECS and MECS.  Background subtraction 
was performed using standard blank field exposures. The average background 
subtracted source 
count rates are reported in Table\,1 for the four observations, together with \
the off--axis angle and number of working MECS. 
The  PHA spectra were rebinned so as to have $>$40 counts in each energy bin and 
reliably adopt a minimum $\chi^2$ technique for model fitting. 
All the bins which were consistent with zero
after background subtraction were rejected. Moreover the MECS spectra were restricted 
to the 1.8--10~keV range. Data from observation B and C were not used for spectral 
analysis purposes owing to poor statistics. 
A constant factor free to vary within a predetermined range was applied in the 
fitting to allow for known normalization differences the LECS and MECS.  

In order to compare our results with previous observations, the spectra were 
first fit with two models: (i) an absorbed power--law, and (ii) an absorbed 
power--law plus blackbody (see Table\,2).
The power--law model gave an unsatisfactory description of the 
spectra with \rchisq\ = $\chi^2$/degrees of freedom (hereafter dof) of 1.6 (347 dof) 
and 1.9 (396) for observations A and D, respectively. 
We note that among simple single component spectral models, the power--law 
gives the lowest reduced $\chi^2$; this is for a photon index $\Gamma$ = 
4.37$\pm$0.03 (A) and 4.55$\pm$0.03 (D; 90\% confidence level 
uncertainties are used throughout the paper). 
The power--law plus blackbody model gave \rchisq\ = 1.28 (344) and 1.07 
(290) for observations A and D, respectively (see Table\,2 for details).  

An F--test shows that the inclusion of the blackbody component is highly 
significant (probability
of $\sim$10$^{-26}$ and $\sim$10$^{-32}$ for obs. A and D, respectively).
The best fit two--component spectra are plotted in Fig.\,1.  
In Fig.\,2 the unfolded energy spectrum for observation A is shown together 
with the contributions of the two spectral components, the power--law and 
the blackbody.  
\begin{table*}
\begin{center}
\caption[]{\BSAX\ phase averaged fit of \src.}
\begin{tabular}{lcccccc}
\hline
Spectral Parameter & Obs. A & Obs. A$^b$ & Obs. D & Obs. D$^b$ & 
Obs. A$^a$+D & Obs. (A+D)$^b$\\
\hline 
 \nh (10$^{22}$\hcm)\,............................... & 1.11$\pm$0.07 & 1.1 $\pm$0.1 & 
0.98$\pm$0.06 & 1.2$\pm$0.1 & 1.12$\pm$0.06 & 1.0$\pm$0.1\\
$\Gamma$\,.............................................................. &   
3.86$\pm$0.06 & 3.8 $\pm$0.3 & 3.58$\pm$0.12 & 4.0$\pm$0.2& 3.95$\pm$0.05 & 3.6$\pm$0.3\\
PL flux (erg s$^{-1}$cm$^{-2}$; 0.5--10 keV)\,........& 7.4$\times10^{-11}$  
& 6.0$\times10^{-11}$ &7.3$\times10^{-11}$ & 8.4$\times10^{-11}$ &7.9$\times10^{-11}$ 
 & 7.7$\times10^{-11}$\\
BB kT (keV)\,\,........................................... &  0.42$\pm$0.02 & 
0.37$\pm$0.02& 0.36$\pm$0.01 & 0.41$\pm$0.04& 0.40$\pm$0.01 &0.36$\pm$0.02\\
BB radius (km @ 1\,kpc)..............................& 1.5$\pm$0.2 & 2.2$\pm$0.5 &
2.1$\pm$0.2 & 1.5$\pm$0.5 & 1.8$\pm$0.2& 2.3$\pm$0.4  \\ 
BB flux (erg s$^{-1}$cm$^{-2}$; 0.5--10 keV)\,........& 2.9$\times10^{-11}$
& 2.6$\times10^{-11}$  &  4.1$\times10^{-11}$ & 2.1$\times10^{-11}$ & 3.0$\times10^{-11}$ & 
2.3$\times10^{-11}$\\
$\chi^2$/dof\,......................................................  &  
440/344 & 214/209  & 309/290 & 203/187  &  455/415  & 283/279 \\
$L_X$ (10$^{34}$ erg s$^{-1}$ @ 1\,kpc; 0.5--10 keV).......& 7.9 & 8.3 &
6.3 & 7.5  & 9.3  & 5.7 \\
\hline
\hline
\end{tabular}
\end{center}
Note --- Fluxes are not corrected for the interstellar absorption. Flux uncertainties 
are about 10\%. The source luminosities were derived by setting N$_H$ = 0. \\
$^a$ Only MECS2 and MECS3 considered.\\
$^b$ Only LECS data considered.
\end{table*}
\begin{table*}
\caption[]{Period history for \src. 90\% uncertainties are reported.}
\begin{center}
\begin{tabular}{llccl}
\hline
Mission &Instrument     & Period  & Date & Reference\\
        &               & (s)     & (year) & \\
\hline
\E&SSS      & 8.68707$\pm$0.00012   & 1979.16 & White \etal 1996  \\
\E&MPC      & 8.68736$\pm$0.0007    & 1979.67 & White \etal 1996  \\
{it EXOSAT}&ME         & 8.68723$\pm$0.00004   & 1984.66 & Israel \etal 1994 \\
\R&PSPC        & 8.68784$\pm$0.00004   & 1993.12 & Hellier 1994      \\
ASCA&GIS          & 8.68791$\pm$0.00015   & 1994.72 & White \etal 1996  \\
\RXTE&PCA           & 8.6881$\pm$0.0002     & 1996.32 & this work  \\
\RXTE&PCA           & 8.688068$\pm$0.000002 & 1996.24 & Wilson \etal 1998 \\
\BSAX&MECS          & 8.68804$\pm$0.00007   & 1997.01 & this work\\
\BSAX&MECS          & 8.6882$\pm$0.0002     & 1997.61 & this work\\
%SAX/MECS         & --                    & 1998.07&  \\ 
\BSAX&MECS          & 8.6883$\pm$0.0001     & 1998.10 & this work\\
\hline
\hline
\end{tabular}
\end{center}
\end{table*}
We also fit the LECS and MECS spectra of observation A with other 
two--component spectral models, such as a power--law with a cut--off 
(\chisq/dof = 453.2/344), two blackbodies (\chisq/dof = 503.1/344) 
and a broken power--law (\chisq/dof = 400.6/344). All {\rc these} models 
gave substantially worse fits than the power--law plus 
blackbody model. Similar results were obtained for the spectrum of observation D. 
Since the blackbody peaks are close to the lower end of the MECS energy range 
we also checked these results by analysing the LECS data only (which cover a 
wider spectral range than the MECS;see Table 2 and Discussion).   

We also checked the stability of the results obtained by rebinning by a factor 
of $\sim$3 the PHA channels in the LECS and MECS spectra. Again all the 
bins which were consistent with zero after background subtraction were 
rejected from the analysis. No significant variation in the spectral parameters 
and uncertainties were found for either models.    
%FIGURE 3
\begin{figure}[hbt]
\centerline{\psfig{figure=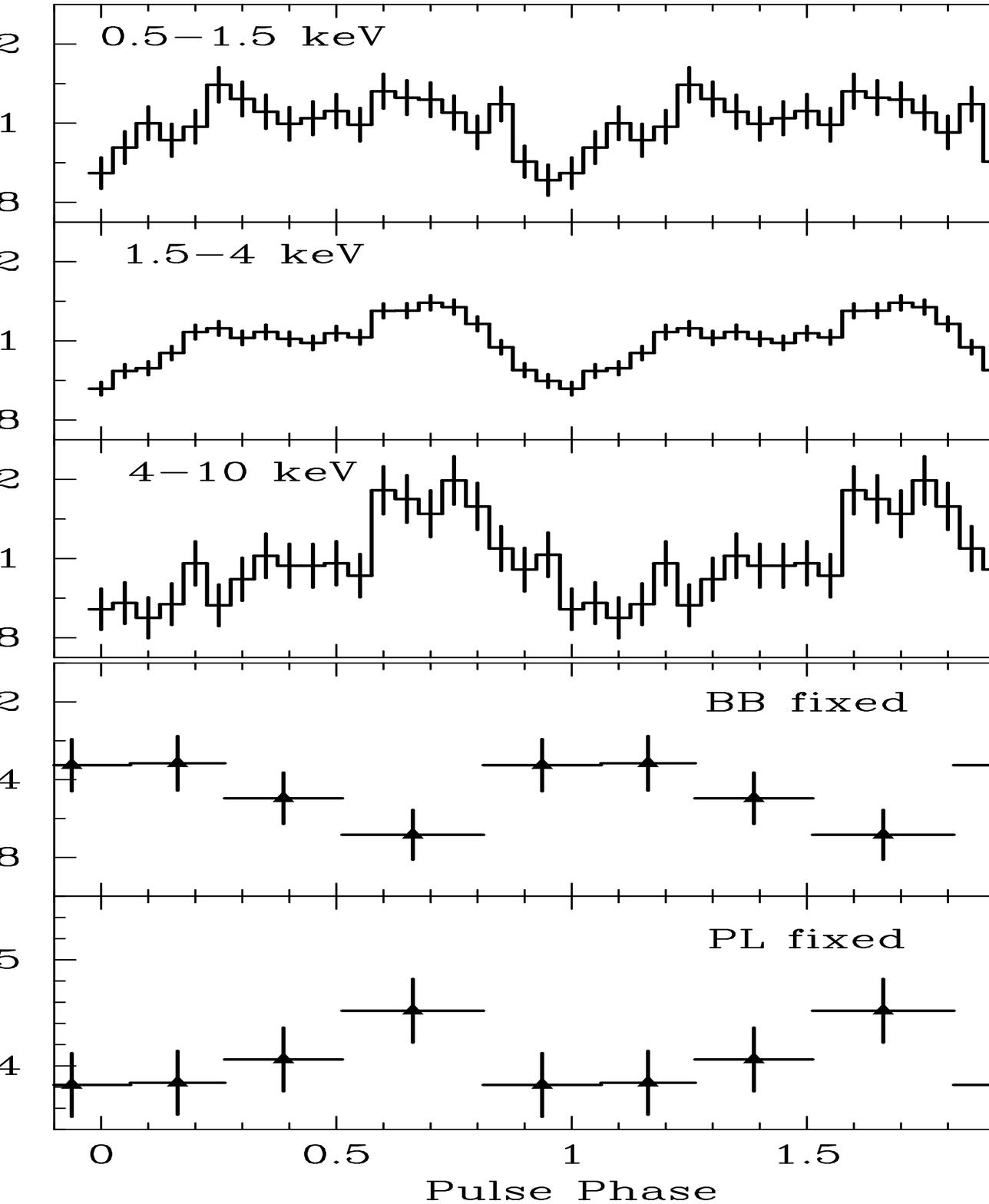,width=6cm,height=12cm}}
\caption{The \src\ MECS light curves for obs. A (first three panels) folded to 
the best period (P=8.68804\,s) in the 0.5--1.5 keV, 1.5--4 keV and 
4--10 keV energy bands. Zero phase was (arbitrarily) chosen to correspond to the 
minimum in the 1.5--4 keV folded light curve. The results of the pulse phase 
spectroscopy of obs. A (last two panels) are also reported for two spectral 
parameters. For clarity two pulse cycles are shown.}
\end{figure}

The data from the {\rc High Pressure Gas Scintillation Proportional Counter (HPGSPC) and 
the Phoswich Detector System (PDS)} did not hold any useful 
information on \src. In fact, due to the large FOVs of these intruments and 
the steep spectrum of \src, the counts above 10 keV were likely dominated 
by the nearby source  RX J0146.9+6121.

\subsection{Pulse timing, folded light curves and phase resolved spectroscopy}

The arrival times of the 0.5--10~keV photons from \src\ 
were corrected to the barycenter of the solar system and 1\,s 
binned light curves accumulated for each observation. 
The average count rates are reported in Table\,1.   

The MECS counts were used to determine the \src\ pulse period. 
The data from observations A, B and D were divided
into 6, 4, and 5 time intervals, respectively and for 
each interval the relative phase of the pulsations was determined. 
These phases were then fit with a linear function giving a best--fit 
period of $8.68804 \pm 0.00007$~s for observation A  (see Table\,3). 
For observation B a period of $8.6882 \pm 0.0002$~s was obtained, while 
during observation C pulsations were not detected 
owing to poor statistics ($\sim$ 11000 photons) to detect such a weak 
($\sim$6\% pulsed fraction) signal.
Finally for observations D we determined a period of 8.6883$\pm$0.0001\,s.
The \BSAX\ pulse period values are plotted in Fig.\,4 together with the 
previous measurements (see Sect.\,3 for the \RXTE\ measurement). The 
background subtracted light curves from observation A, folded at the best 
period in different energy ranges (Fig.\,3; first three panels) show a 
double--peaked profile (see also White \etal 1996).  
The pulsed fraction (semiamplitude of modulation divided by the mean source 
count rate) was 7.1$\pm$1.5\%, 7.5$\pm$0.5\% and 13$\pm$2\% in the 
0.5--1.5 keV, 1.5--4.0 keV and 4.0--10 keV energy bands, respectively. 
Similar results are obtained for observations B and D.  

%
%FIGURE 4
\begin{figure}[hbt]
\centerline{\psfig{figure=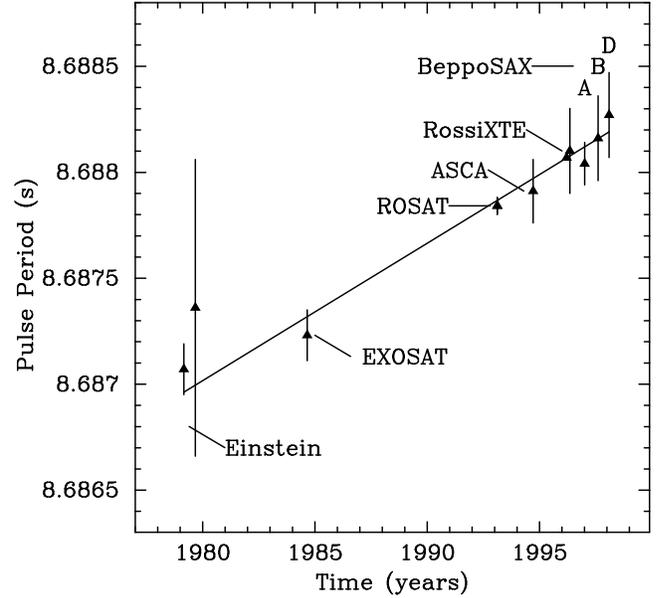,width=6cm,height=8cm}}
\caption{Pulse period history of \src\ as a function of time. 1$\sigma$ 
uncertainties have been used.}
\end{figure}

Flux variations on several timescales are usually displayed   
by high accretion rate X--ray pulsars; we found  no short or long--term 
variations in any of the \BSAX\ accumulated light curves. 

The upper two panels of Fig.\,5 show the pulsed fraction versus energy during 
observation A for the first two harmonics. These values were obtained by 
fitting the corrisponding light curves with two sinusoidal functions. The first 
harmonic shows a nearly constant value ($\sim$ 6\%) up to 4\,keV, while at  
higher energies increases to about 15\%. A constant value of 5--7\% 
is inferred for the second harmonic. We calculated also the root mean square 
variability (rms; defined as $\sqrt{V_{obs}-V_{exp}}$ devided by the mean source 
count rate, where $V_{obs}$ and $V_{exp}$ are the observed and expected variance, 
respectively) of the folded light curve at different energies (third panel).  
%FIGURE 7
\begin{figure}[tbh]
\centerline{\psfig{figure=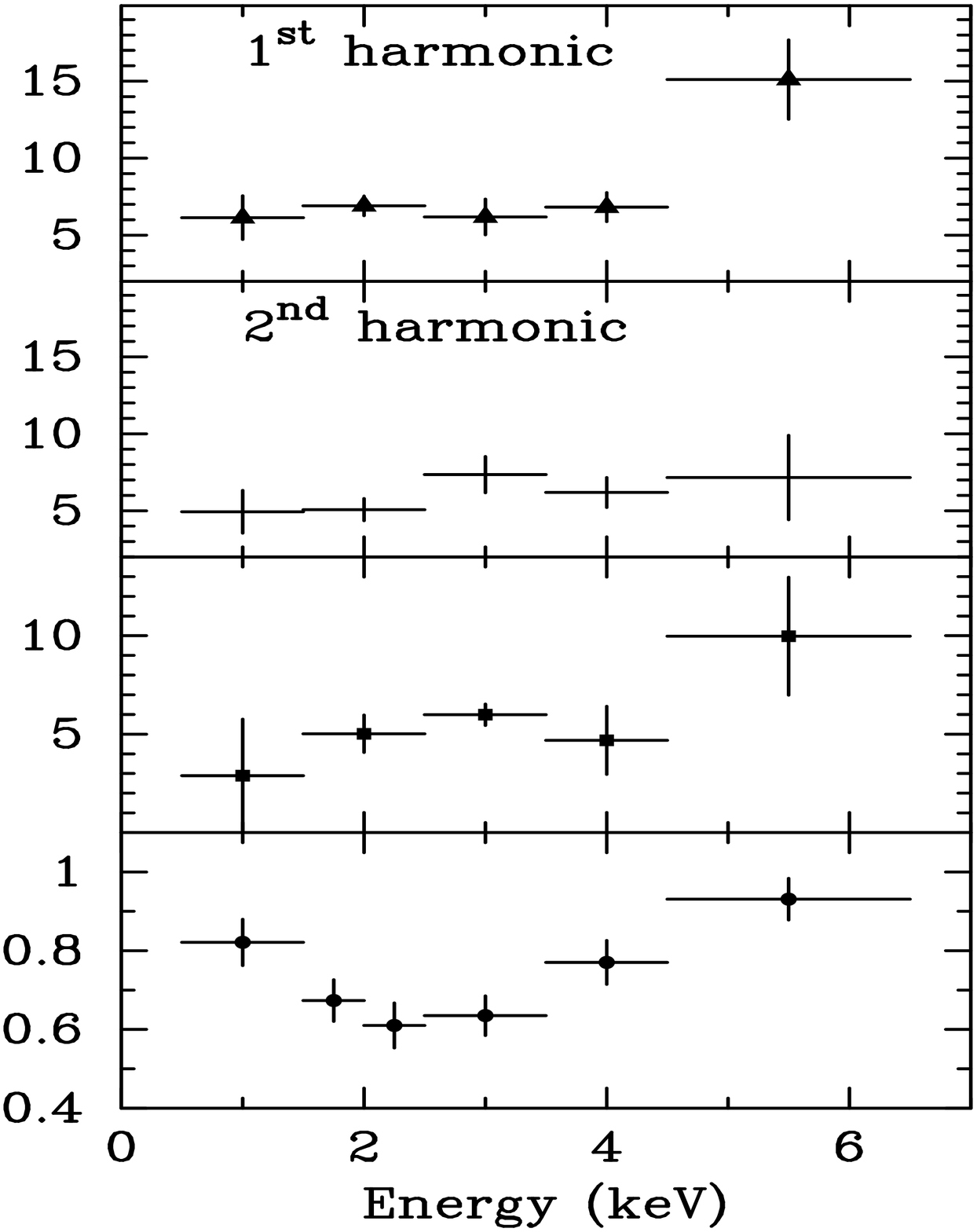,width=6cm,height=11cm}}
\caption{Pulsed fraction of the first two harmonics of \src\ as a function of 
the energy for observation A (upper two panels), together with the rms and the 
power--law to total flux ratio (lower two panels). See text for details.}
\end{figure}
Finally the ratio between the power--law and the total flux for the blackbody plus 
power--law spectral model is shown (lowest panel). From the comparison of these 
quantites we can infer that: (i) since the rms variability of the folded light 
curves is $\pi^{-1/2}$ times the geometric sum of the amplitudes in all available 
harmonics (mainly the 1$^{st}$ and 2$^{nd}$ in the present case), it is apparent 
that the behaviour is consistent with the derived amplitudes of the 1$^{st}$ and 
2$^{nd}$ harmonics; (ii) there is evidence for an increase of the pulsed fraction 
($\sim$ 2$\sigma$ confidence level) at energies above 5 keV; (iii) there is not a 
simple relationship between the pulsed fraction and the flux in either of the two 
spectral component adopted in our analysis. 

A set of four phase--resolved spectra (phase boundaries 0.06, 0.26, 0.5, 0.78) 
were accumulated for observation A (i.e. the observation with the highest 
number of source counts).  
These were then fit with the power--law plus blackbody model described 
in Sect.\,2.1, with N$_H$ fixed at the phase--averaged best--fit value. 
Initially, the blackbody temperature was fixed at its  phase--averaged 
best--fit value and only the power--law parameters and blackbody normalisation 
were allowed to vary. The fits were then repeated with $\Gamma$ fixed 
and blackbody parameters and power--law normalisation free {\rc (see Fig.\,3). 
%The fits obtained were in both cases acceptable 
No significant changes were detected for $\Gamma$ and $kT$ as a function 
of the pulse phase. Similar results were obtained for the fluxes of the two 
spectral component.}      
 
\section{\RXTE\ Observation}

The \src\ position was included in a \RXTE\ observation pointed at 
the nearby high--mass X--ray binary pulsar RX J0146.9+6121 
(from 1996 March 28 11:16:48 to 22:09:20; $\sim$20~ks of effective 
exposure time). The results presented here are based on data 
collected in the so called ``Good Xenon" operating 
mode with the Proportional Counter Array (PCA, Jahoda et al. 
1996). The PCA consists of 5 
proportional counters operating in the 2--60 keV range, 
with a total effective area of approximately 7000~cm$^{2}$  and a 
field of view, defined by passive collimators, of $\sim1\deg$   
FWHM. The data consist of the time of arrival and the pulse height 
of each count. In order to minimise the contamination from 
RX\,J0146.9+6121 (a much harder spectrum X--ray pulsar), we considered 
only photons in the 2--4 keV energy 
interval. In order to reduce the background, only the counts detected in the 
first Xenon layer of each counter were used. 
%FIGURE 6
\begin{figure}[tbh]
\centerline{\psfig{figure=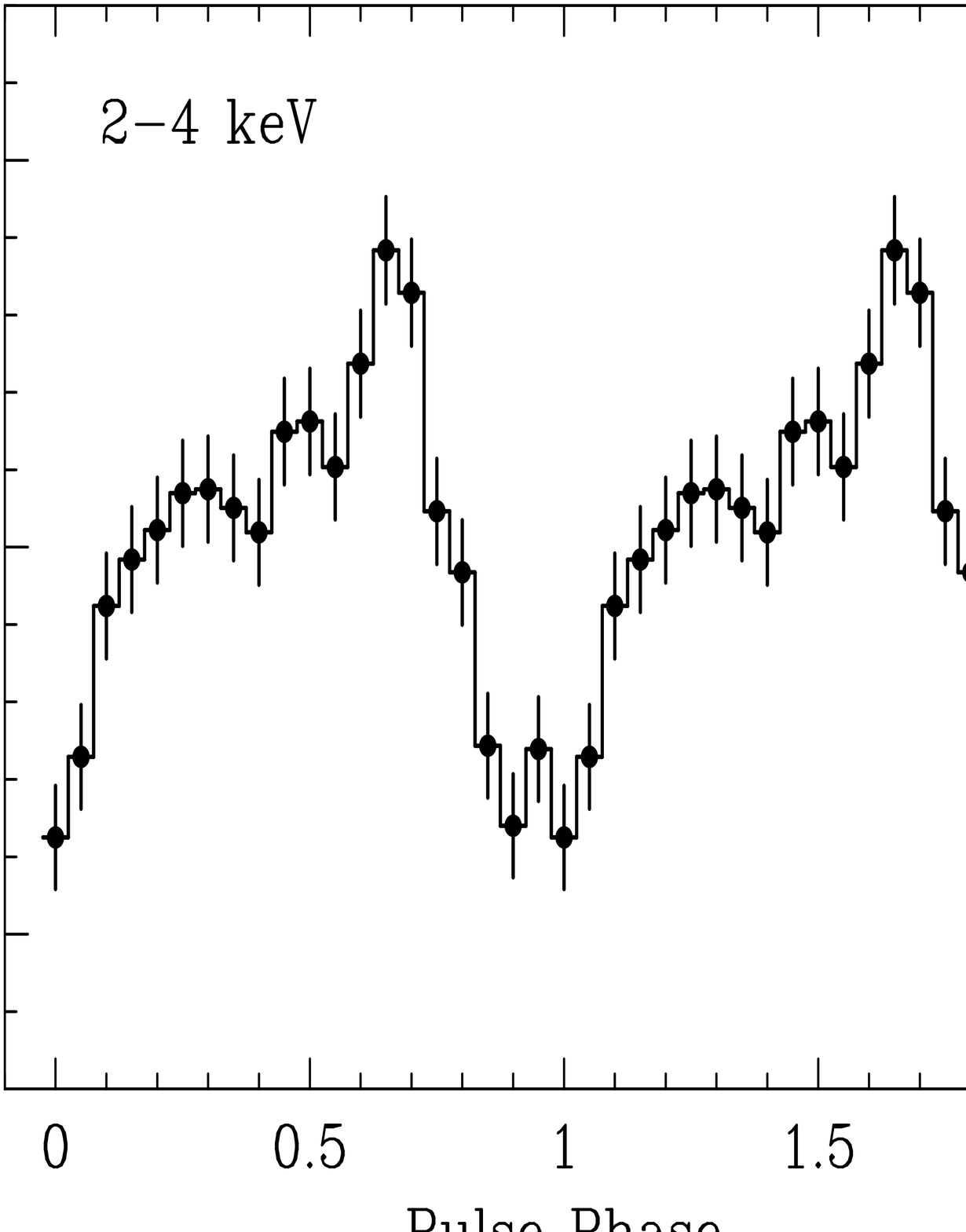,width=6cm,height=4cm}}
\caption{The \src\ \RXTE\ 2--4 keV light curve folded to 
the best fit period (P=8.6881\,s).}
\end{figure}

We obtained for \src\ a best pulse period of 8.6881$\pm$0.0002\,s 
(see Table\,3 and Fig.\,4).  
Figure\,6 shows the corresponding folded light curve. 
Despite the uncertainties in the background subtraction (the contribution from 
RX\,J0146.9+6121 is difficult to estimate), it is apparent that the \RXTE/PCA 
pulse profile of \src\ is similar to that observed with \BSAX. 
The \RXTE\ value we obtained is consistent to within the statistical uncertainties 
with that of Wilson \etal (1998; energy range 3.7--9.2 keV). The somewhat different 
pulse shape is likely due to the different energy band used.   

\section{Discussion}

Several models have been proposed in order to explain 
the  nature of the ``anomalous'' X--ray pulsars. 
Mereghetti \& Stella (1995) proposed that 
these sources form a homogeneous subclass of accreting neutron stars, perhaps 
members of low mass X--ray binaries (LMXBs), which are characterized by lower 
luminosities  
($10^{35}-10^{36}$~erg~s$^{-1}$) and higher magnetic fields ($B\sim10^{11}$~G)
than classical LMXBs. However the lack of evidence for a binary nature from 
any of these systems (Mereghetti \etal 1998; Wilson \etal 1998) 
argues in favor of models in which the X--ray emission originates from a compact 
object that is not in an interacting binary system. 

Van Paradijs \etal (1995) proposed that AXPs are young population I objects, which 
originate from the evolution of short orbital period High Mass X--ray Binaries (HMXBs), 
following the expansion of the massive 
star and  the onset of unstable Roche Lobe overflow, before central hydrogen is 
exhausted (see also Cannon \etal 1992). The resulting 
common--envelope phase should {\rc cause the neutron star to spiral--in and 
disrupt the companion star after the so--called
Thorne--Z\.ytkov stage.}  Therefore, these sources should consist of an 
isolated neutron star accreting matter from a residual disk 
with a mass in the $10^{-3}-1$~M$_{\odot}$ range. The blackbody component 
in \src\ and, likewise, other AXPs has been interpreted as evidence for 
quasi--spherical accretion onto an isolated neutron star formed after common envelope 
and spiral--in of a massive short--period (P$_{orb}$$\leq$ 1\,yr) X--ray binary. 
In this case the remains of the envelope of the massive star might produce two 
different types of matter inflow: 
a spherical accretion component with low specific angular momentum giving rise 
to the blackbody component and a high specific angular momentum component which 
forms an accretion disk and is likely responsible for the power--law emission 
(Ghosh \etal 1997).        
Although some problems 
remain open in this scenario, we note that a post common--envelope 
evolutionary scenario for 4U0142+614 is also suggested by the 
inference that about half of the X--ray absorbing material is close to the 
source (White \etal 1996). 
In this context it is interesting to note that independent evidence supports
the view that the binary X--ray pulsars 4U1626--67 and HD49798 were formed 
following a common envelope and spiral--in evolutionary phase (see Angelini 
\etal 1995 and Israel \etal 1997 and references therein). The different
properties of the companion stars in these two cases may reflect the fact 
that unstable mass transfer set in at different evolutionary phases in the 
nuclear evolution of their progenitors (in turn, reflecting different 
initial masses and orbital periods; see also Ghosh \etal 1997).  

Thompson \& Duncan (1993, 1996) proposed that the AXPs are ``magnetars'', 
neutron stars with a superstrong  magnetic field ($\sim10^{14-15}$ G).
This proposal is supported by the similarity in the pulse periods 
(8.05\,, 7.5\,s and 5.16\,s) and of \.P values measured in the Soft 
$\gamma$--ray Repeaters SGR0520--66 , SGR1806-20 and SGR1900+14, 
respectively  (Kouveliotou \etal 1998; Hurley \etal 1998).   
If this connection proved correct, AXPs, would be quiescent soft $\gamma$--ray 
repeaters. Heyl \& Hernquist (1997) argued that their 
emission may be powered by the cooling of the core through a 
strongly magnetised  envelope of matter (made up mainly by hydrogen and helium).  
Pulsations would originate from a temperature gradient on the surface of the star. 
Moreover Heyl \& Hernquist (1998) showed that in this scenario, spin--down 
irregularities, observed in the period history of two ``anomalous'' X--ray 
pulsars (1E\,2259+58 and 1E\,1048--59), may be simply accounted for with glitches 
like those observed in young radio pulsars.

The 0.5--10 keV spectrum of \src\ is well modelled by the sum of an absorbed steep 
power--law and a low--energy blackbody. The latter component was introduced 
by White \etal (1996) based on the ASCA data; the corresponding blackbody 
radius  and flux were $\sim2.4\pm0.3$\,km (at 1 kpc) 
and $\sim$40\% of the total, respectively.  {\rc Neither the power--law photon 
index nor the blackbody temperature show} evidence 
of changes across different \BSAX\ observations.  We found marginal evidence 
(at about 2$\sigma$ confidence level) for changes relative to previous 
observations. A comparison of the ASCA and \BSAX\ results show that: 
(i) the power--law photon index increases by about 4\% (observations A and D); 
(ii) the blackbody radius decreased by about 37\% (A) and 12\% (D); (iii) 
the blackbody flux decreased to $\sim$30\% (A) and $\sim$35\% (D) of the 
total flux therefore showing a $\sim$10\%--5\% decrease with respect to that of the 
ASCA observation; (iv) the 0.5--10 keV total flux decreased of about 15--12\%. 
 
By fixing the parameters N$_H$, $kT$ and $\Gamma$ to the values inferred by ASCA, 
the \BSAX\ observation A spectrum gives a \rchisq/dof of 1.9/348. 
In this case the blackbody component accounts for $\sim$30\% of the total flux 
($\sim$10\% lower than that of ASCA). 
As an additional test we also merged the data for observation A (MECS 2 and 3 
only) and D and fit them with the power--law plus blackbody model (see Table\,2). 
While a small change in the spectral parameters is found relative to observations 
A and D, separately, the blackbody flux is $\sim$30\% of the total. 

To remove possible effects introduced by the vicinity of the lower end of the energy 
range of the MECS and the peak of the blackbody component, 
we also fitted the spectrum of observations A, D and A+D by using only the data 
as accumulated by the LECS, the energy band of which uninterruptely covers the 
range 0.5--9.0 keV 
and allows a better fit in the energy interval (1--4 keV) where the power--law 
and the blackbody components overlap. Again we obtained results similar to the 
LECS+MECS case, but with a larger uncertainty. 
All these results point to a marginal variation of the spectral 
parameters of \src, if any.

The \BSAX\ data {\rc suggest} that the pulsed 
fraction for energies above 4\,keV decreased (from 25\%$\pm$5\% with ASCA to 
13\%$\pm$2\% with \BSAX; 90\% uncertainties). For these energies the counts are  
almost entirely dominated by the power--law component (in the spectral model assumed). 

The pulse periods measured by \BSAX\ show that \src\ has continued its secular 
spin--down during 1996--1998 (see Fig.\,4). The period derivative inferred 
from the \BSAX\ observations alone ($\sim$6.0$\pm^7_{5.1}$
$\times$10$^{-12}$ s s$^{-1}$) is consistent with the average spin--down rate 
($\sim$2$\times$10$^{-12}$ s s$^{-1}$) inferred over the 19\,yr span of the 
historical dataset (note that in the period list we did not include the low 
significance detections inferred  on 1985 November 11 and December 11 with EXOSAT 
ME, and on 1991 February 13 with ROSAT HRI; Israel \etal 1994).  
The inferred \.P for \src\ is consistent with the uncertainties of all pulse 
period measurements except, perhaps, one (the 1979 \E\ SSS one; see Fig.\,4) 
implying that, so far, no ``glitches'' have been observed yet.  

\begin{acknowledgements} 
We would like to thank Giancarlo Cusumano for providing the MECS off--axis 
matrices and the \BSAX\ Mission Planning for their constant help. The authors 
also thank K. Long the comment of which helped to improve an earlier version 
of this paper. This work was partially supported through ASI grants.
\end{acknowledgements}

\end{document}